\documentclass[prl,aps,twocolumn,superscriptaddress]{revtex4}

\usepackage{bm}
\usepackage{epsfig}
\usepackage{amsmath}
\usepackage{amssymb}

\usepackage[usenames]{color}

\begin{document}

\definecolor{Blue}{rgb}{0,0.0,1.0}
\definecolor{Red}{rgb}{1.0,0.0,0.0}
\newcommand{\comment}[1]{\textcolor{Blue}{#1}}
\newcommand{\change}[1]{\textcolor{Red}{#1}}

\title{Preparation and relaxation of very stable glassy states of a simulated liquid}

\author{Robert L. Jack}
\affiliation{Department of Physics, University of Bath, Bath BA2 7AY, United Kingdom}

\author{Lester O. Hedges}
\affiliation{Lawrence Berkeley National Laboratory, Berkeley CA, 94720}

\author{Juan P. Garrahan}
\affiliation{School of Physics and Astronomy, University of Nottingham, Nottingham, NG7 2RD, United Kingdom}

\author{David Chandler}
\affiliation{Department of Chemistry, University of California, Berkeley CA, 94720}

\begin{abstract}
We prepare metastable glassy states in a model glass-former made of Lennard-Jones particles
by sampling biased ensembles of trajectories with low dynamical activity.  These trajectories form an inactive dynamical phase whose `fast' vibrational degrees of freedom are maintained at
thermal equilibrium by contact with a heat bath, while the `slow' structural degrees of freedom are located
in deep valleys of the energy landscape.   
We examine the relaxation to equilibrium and the vibrational properties of these metastable states.  
The glassy states we prepare by our trajectory sampling method are very stable to thermal fluctuations and also more mechanically rigid than low-temperature equilibrated configurations.
\end{abstract}

\maketitle

\newcommand{\tobs}{t_\mathrm{obs}}
\newcommand{\ee}{\mathrm{e}}

\newcommand{\msd}{\langle r^2(t)\rangle}

As a supercooled liquid is cooled towards its glass transition, its viscosity increases dramatically while
its structure changes only subtly~\cite{ediger96,deb-still01,cavagna,ARPC,berthier-review}.   
Thus, 
different fluid states with similar structures may have relaxation times that differ by many orders of magnitude.  
In this report, we focus on fluid configurations that relax especially slowly.
We do so with a field $s$ that suppresses trajectories with appreciable particle motion~\cite{merolle-jack,lecomte07,garrahan-fred,hedges09,jack10-rom,elmatad10-pnas}. 
It is this field that controls a dynamical or {\em space-time} phase transition~\cite{lecomte07,garrahan-fred,hedges09} in glass forming liquids, a transition between {\em active} fluid states and {\em inactive} states where structural relaxation may be completely arrested.  

We consider a binary mixture of spherical particles which supports both active and inactive states.  The structure of the inactive state differs subtly from the active one, and these differences render the inactive state extraordinarily stable.  Thus, while the field $s$ biases the \emph{dynamics} of the system, the fluid responds by changing its \emph{structure}, so as to arrive in long-lived metastable states.  
We find that these states are located in (or near~\cite{Kurchan96}) deep valleys of the energy landscape~\cite{Heuer08,cavagna}.  
The relationships between long-lived metastable states and glassy behaviour have been discussed 
extensively~\cite{cavagna,tap,BK01,heuer03,pts08,KL11}.  
However, even the definition of a metastable
state requires a dynamical construction that accounts for its lifetime~\cite{BK01,KL11}, while the energy landscape
is a purely static object. Since the field $s$ 
couples directly to the dynamical evolution of the system, we find that it is a powerful new tool for
analysing long-lived metastable states.

The model we study is the Lennard-Jones (LJ) mixture of Kob and Andersen~(KA)~\cite{ka95a}. 
There are $N$ particles in the system, of which $N_{\rm A}=0.8N$ are of type A
and $N_{\rm B}=0.2N$ are of type B.
The unit of length is the diameter $\sigma$ 
of the type A particles, and we set the LJ energy for AA interactions to be $\epsilon=1$.  
All particles have mass $m$ and we take Boltzmann's constant $k_{\rm B}=1$.  
To facilitate sampling of the $s$-ensemble, we consider a small system of $N=150$ particles
in a box of size $(5\sigma)^3$ with periodic boundaries, as in~\cite{hedges09}.

The system is coupled to a heat bath so its dynamical evolution is stochastic.
We consider both Newtonian dynamics coupled to a thermostat, and a Monte Carlo (MC) dynamical
scheme.
Both methods give similar results,
both at equilibrium~\cite{berthier-kob07} and in the $s$-ensemble~\cite{hedges09}.
We use $x=(\bm{r}_1,\bm{r}_2,\dots,\bm{r}_N)$ to represent the positions of all particles in the system.  
We consider ensembles of trajectories (`$s$-ensembles') based on large deviations~\cite{lecomte07} of the dynamical activity.  Within the $s$-ensemble,
trajectories have length $\tobs$ and the probability of a trajectory $x(t)$ is 
\begin{equation}
\mathrm{Prob}[x(t)|s] = \mathrm{Prob}[x(t)|0]\frac{\ee^{-sK[x(t)]}}{{\cal Z}(s)},
\label{equ:s-ens}
\end{equation}
where $\mathrm{Prob}[x(t)|0]$ is the probability of the trajectory $x(t)$ at equilibrium
and ${\cal Z}(s)$ is a normalisation
factor.  The dynamical activity $K$ measures the amount of motion that takes place in a trajectory, and is defined by
$K=\Delta t\sum_{i=1}^{N_{\rm A}} \sum_{j=0}^{M-1} |\bm{r}_i(t_j+\Delta t) - \bm{r}_i(t_j)|^2$ where the $t_j=j\Delta t$ are equally spaced times along
the trajectory, $M=\tobs/\Delta t$, and the index $i$ runs over all particles of type A.  The method exploits the idea that since
the most striking glassy properties are dynamical in nature~\cite{ARPC,berthier-review}, the
dynamical activity is a natural order parameter for the glass transition~\cite{gc02-prl}.
We sampled these ensembles using transition path sampling~\cite{tps-annrev02,hedges09}.

\begin{figure}
\includegraphics[width=7.5cm]{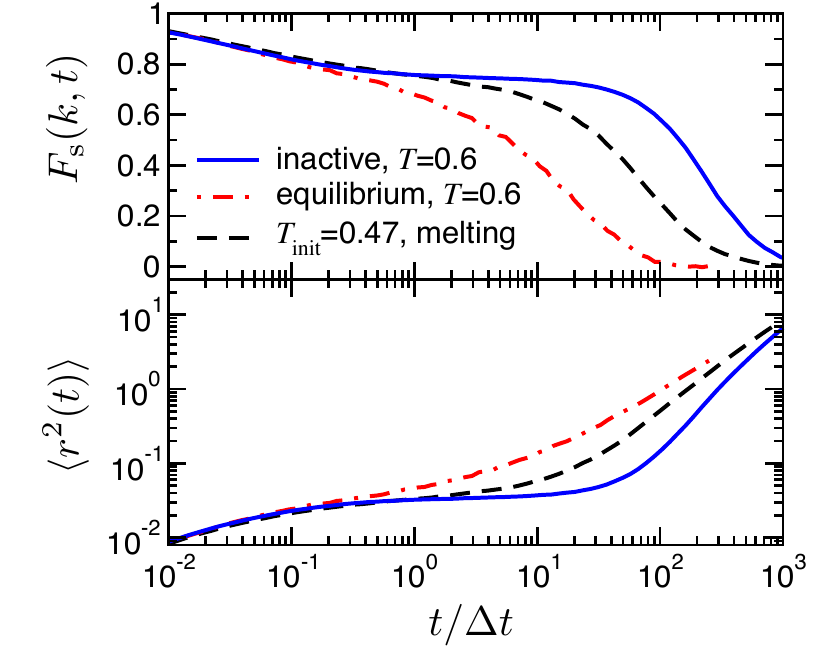}
\caption{
Self-intermediate scattering function, $F_\mathrm{s}(k,t)=N_{\rm A}^{-1} \langle \sum_{i=1}^{N_{\rm A}} \exp{(-{\rm i}\bm{k}\cdot[\bm{r}_i(t)-\bm{r}_i(0)])}\rangle$,  and mean-squared displacement, $\msd$, from simulations at $T=0.6$.  We show time-dependent expectation values
evaluated with equilibrated initial conditions at $T=0.6$ (dot-dashed); from an inactive $s$-ensemble at $T=0.6$ (full line, see the main text for details); 
and from equilibrated initial conditions at $T_\mathrm{init}=0.47$ (dashed line).
In the definition of $F_\mathrm{s}(k,t)$, the sum runs over all particles of type A and $k=|\bm{k}|=7.251/\sigma$ corresponds to the first peak of the structure factor.
}
\label{fig:melt-r2}
\end{figure}

We focus on \emph{inactive configurations} taken from the inactive state in
the $s$-ensemble, and we compare them with thermally-equilibrated configurations.  
To assess the stability of different configurations, we used them
as initial conditions for simulations with MC dynamics, implemented as in~\cite{berthier-kob07,hedges09}. 
All simulations are run at temperature $T=0.6$, and no biasing field
$s$ was applied.
Results are shown in Fig.~\ref{fig:melt-r2}, where we show the mean
square displacement of the type A particles, $\msd$, and also their self-intermediate scattering function,
$F_\mathrm{s}(k,t)$.
We use these simulations
to model the `melting' of the inactive state, and we compare this process with the heating of a 
supercooled liquid state from one temperature
to another (see also the recent experiments in~\cite{ediger07}).
In our MC simulations the unit of time is $\Delta t$, defined such that the
diffusion constant in the limit of low density is $D_0=\sigma^2/\Delta t$~\cite{hedges09}.
For simulations with Newtonian dynamics, we take $\Delta t=1.92\sqrt{m\sigma^2/\epsilon}$ which allows quantitative
comparison with MC results.
Inactive configurations were obtained from the mid-point ($t=\tobs/2$) of
trajectories $x(t)$, taken from an $s$-ensemble with MC dynamics at
$T=0.6$, $\tobs=150\Delta t$ and $s=0.0725/(\sigma^2\Delta t)$.
This $s$-ensemble
is in the inactive state: we have considered other ensembles from this state and their behaviour is qualitatively similar.

For simulations with inactive initial configurations, $\msd$ shows a plateau, with the system remaining stable for 
at least $50\Delta t$ before the particles diffuse away from their initial positions.   We conclude that the inactive
configurations are localized in metastable states,
and must overcome significant free energy barriers before they relax to equilibrium.   
Comparing initial conditions from the inactive phase with 
equilibrated fluid configurations from $T=0.47$, we see that these fluid states are less stable, and relax
more quickly to equilibrium.
While steady state simulations at equilibrium and in the $s$-ensemble are similar for both MC and Newtonian dynamics,
melting and heating processes do depend significantly on the dynamics used in our simulations.  MC dynamics
approximate the overdamped limit of strong coupling to a heat bath, and are
convenient for demonstrating the metastability of the inactive phase, as in Fig.~\ref{fig:melt-r2}.

\begin{figure}
\includegraphics[width=8cm]{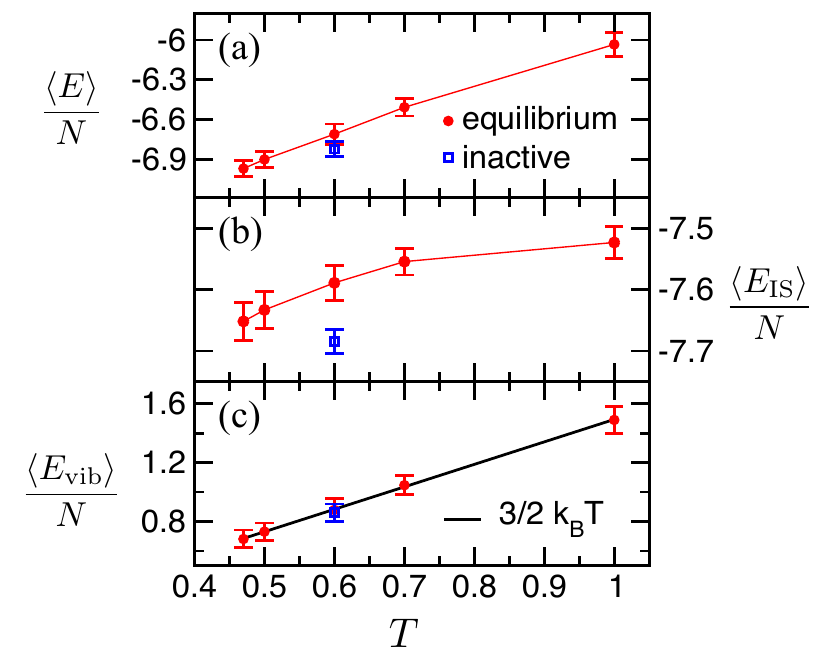}
\caption{
  Average energy $\langle E\rangle$, average inherent structure energy
  $\langle E_{\rm IS}\rangle$ and average vibrational energy $\langle E_\mathrm{vib}\rangle$, for equilibrium states at various
  temperatures, and for inactive configurations.  Error bars 
  show the size of sample-to-sample fluctuations for these small systems; numerical
  uncertainties are much smaller than these error bars.  In (c), the solid line
  is the result for harmonic vibrations, $\langle E_\mathrm{vib}\rangle=\frac32 N k_{\rm B} T$.
  (The ensemble of inactive configurations is the same as that in Fig.~\ref{fig:melt-r2}.)
}
\label{fig:en-ave}
\end{figure}

In Fig.~\ref{fig:en-ave}(a), we show the average energies $\langle E \rangle$ for equilibrated states at various temperatures,
and for the inactive configurations.  
The energy of the inactive state is lower than the equilibrated state at the same temperature, but this difference is small compared
to the variation in energy between different equilibrated states.  Given that the inactive configurations are much more
stable than the thermally-equilibrated ones, their relatively large energy may seem surprising.

To understand this result, we consider inherent structures (ISs)~\cite{still-web84}, 
obtained by using a conjugate gradient method to find the 
`nearest' energy minimum to any configuration.  
The energy of configuration $x$ is $E(x) = E_\mathrm{IS}(x) + E_\mathrm{vib}(x)$ where $E_\mathrm{IS}(x)$ is the energy
of the inherent structure associated with $x$ and 
we loosely identify $E_\mathrm{vib}(x)$ with `vibrations' around the IS.  
Fig.~\ref{fig:en-ave}(b,c) shows the averages of $E_\mathrm{IS}$ and $E_\mathrm{vib}$. 
The inactive configurations have IS energies that are lower than any of the equilibrated systems we considered.
In computer simulations, the KA mixture has been equilibrated at temperatures as low as $T=0.42$~\cite{berthier2004}. 
The average inherent structure energy in the inactive state appears to be 
consistent with that of equilibrated states near to or below this temperature.
Making the simple approximation of thermally-equilibrated harmonic vibrations about the IS positions, we predict 
$\langle E_\mathrm{vib} \rangle = \frac32 Nk_\mathrm{B}T$, consistent with the data for both
both thermally-equilibrated and inactive states [see Fig.~\ref{fig:en-ave}(c)].  

Thus, we attribute the stability of the inactive configurations (Fig.~\ref{fig:melt-r2}) to their low inherent structure energies.  
This link is consistent with studies of the energy landscape at equilibrium, 
although there is also evidence that slow particle motion is correlated not just with deep minima
but also with saddles that have few unstable directions~\cite{cavagna,Heuer08,saddles}.
Comparing
active (equilibrated) and inactive configurations at $T=0.6$, we see from Fig.~\ref{fig:en-ave}(b)
that the biasing field $s$ has a strong effect on the IS degrees
of freedom, while the vibrational degrees of freedom remain close to equilibrium at temperature $T$.  
Thus, for the relatively small value of $s$ that we are considering, 
it appears that the probability of finding a configuration $x$ in the inactive $s$-ensemble is approximately
\begin{equation}
P(x|s) \propto {\cal P}(x_\mathrm{IS}|s) {\ee^{-E_\mathrm{vib}(x)/T}},
\end{equation}
where ${\cal P}(x_\mathrm{IS}|s)$ is an $s$-dependent statistical weight associated with the inherent structure $x_\mathrm{IS}$, while
the Boltzmann factor 
on the right hand side indicates that the vibrational degrees of freedom are close to equilibrium at the bath temperature. 
At equilibrium, one has ${\cal P}(x_\mathrm{IS}|0)=\ee^{-E_\mathrm{IS}(x)/T}$ but Fig.~\ref{fig:en-ave}(b) shows that 
 ${\cal P}(x_\mathrm{IS}|s)$ is dominated by ISs that are much lower in energy than those found at equilibrium.

\begin{figure}
\includegraphics[width=7cm]{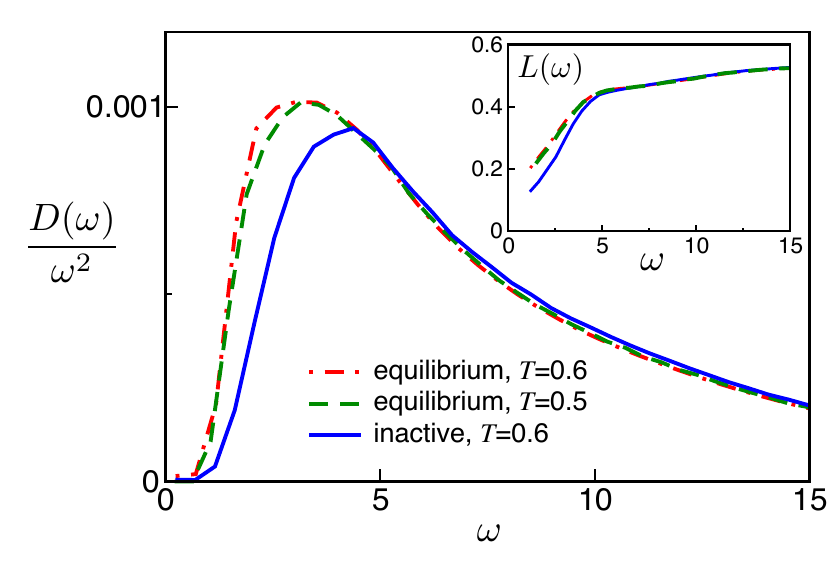}
\caption{
Vibrational density of states $D(\omega)$ (scaled by $\omega^2$) for equilibrium states at $T=0.6$ (dot-dashed) 
and $T=0.5$ (dashed), and inactive states at $T=0.6$ (full line).  Note the relative absence of low frequency modes in the inactive state.  
The inactive data are taken from an $s$-ensemble with Newtonian dynamics and
$\tobs=600\Delta t$ sampled at $s=0.009/(\sigma^2\Delta t)$, near to space-time phase coexistence,
but restricted to $K/(N\tobs\sigma^2)<0.03$~\cite{hedges09}.  Configurations were taken
from all times throughout these trajectories.  This $s$-ensemble was chosen to optimise statistics for $D(\omega)$:
results for the inactive configurations considered in Fig.~\ref{fig:melt-r2} are similar.
The inset shows the participation ratio $L(\omega)$.
}
\label{fig:mode}
\end{figure}

We have calculated the vibrational
densities of states for these states by expanding the energy $E(x)$ around the IS and diagonalizing the Hessian
matrix to obtain (dimensionless) eigenfrequencies $\omega_\alpha$ and eigenvectors $\vec{e}_\alpha$. 
The density of states $D(\omega)$ is the distribution of eigenfrequencies: eigenvectors with small $\omega$ are `soft directions'
on the energy landscape, which may be correlated with the motion of particles during structural 
relaxation~\cite{harrowell,wyart07}.
Fig.~\ref{fig:mode} shows that inactive configurations have fewer soft directions than configurations from thermal equilibrium:
in this sense, the inactive state is more rigid than the thermally equilibrated states.

We also show the participation ratio~\cite{laird}, 
$L(\omega) \equiv \left\langle1/[N \sum_i (\bm{e}^i_\alpha \cdot \bm{e}^i_\alpha)^2]\right\rangle$, where the sum runs over
all particles, the vector $\bm{e}^i_\alpha$ 
contains the components of $\vec{e}_\alpha$ associated with particle $i$, and the average is over modes with frequency 
$\omega_\alpha=\omega$, from all relevant configurations.
In all cases,  $L(\omega)$ decreases for small $\omega$, indicating
that the soft modes are
localized on a relatively small number of particles.  Thus, while the inactive states have fewer soft directions
and hence smaller vibrational fluctuations, the nature of the modes themselves appears similar between
active and inactive states.%

\begin{figure}
\includegraphics[width=7.5cm]{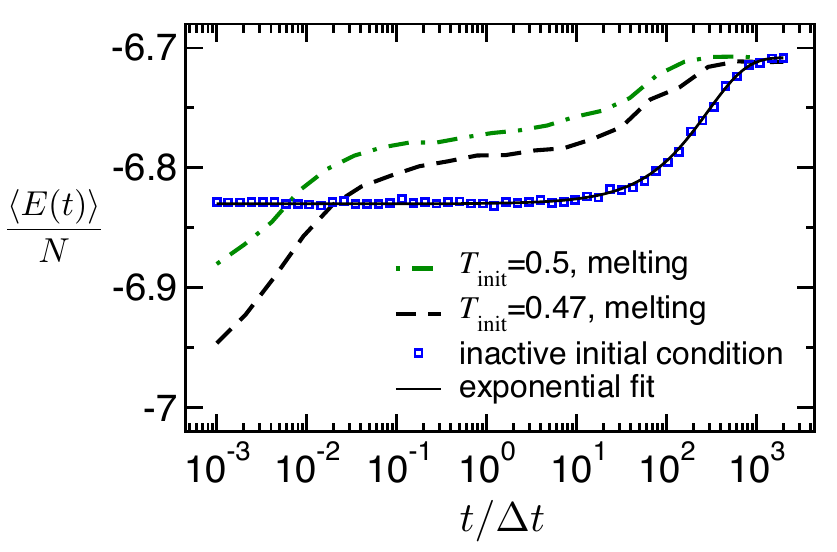}
\caption{
  Time-dependent energy in `melting' simulations at $T=0.6$.  For low-temperature equilibrated initial
  conditions, energy flows into the system in two stages, corresponding to fast ($t\lesssim0.1\Delta t$) and slow
  ($t\gtrsim\Delta t$)
  relaxation.  For inactive initial conditions, there is only a single stage.  The solid black
  line is an exponential fit with characteristic time $290 \Delta t$.
}
\label{fig:en-melt}
\end{figure}

In Fig.~\ref{fig:en-melt}, we show the time evolution of the energy 
for the `melting' simulations discussed above (recall Fig.~\ref{fig:melt-r2}).  
On taking an equilibrated configuration from $T=0.47$ and running MC dynamics at temperature $T=0.6$, 
energy flows into the system in two distinct stages:
the vibrational degrees of freedom respond quickly to the change in temperature while 
the structural degrees of freedom respond more slowly.  
On the other hand, on taking an inactive configuration and running MC dynamics at $T=0.6$, 
the fast degrees of freedom in the inactive state are already close to equilibrium  
and there is no initial stage of relaxation.  
The system remains localised in the metastable inactive state until it finally relaxes back to equilibrium,
with an approximately exponential time-dependence.  

It is natural to ask what structural features of the inactive configurations are responsible for their low IS energies.  
As in~\cite{hedges09}, we exclude crystalline states from the $s$-ensembles
we consider, since we are specifically interested in amorphous glassy states.  
Performing a common neighbour analysis~\cite{honey-cna,ka-cna}, we find that inherent structures from the inactive state
are slightly richer in the `155' environment than their equilibrated counterparts. The 155 environment is associated 
with icosahedral co-ordination~\cite{ka-cna}.  However, the differences are 
subtle and sample-to-sample fluctuations large~\cite{hedges09}: 
we did not find a specific structural motif to which we can attribute the stability
of the inactive configurations.   

We end with a general discussion of the role of metastable states in the $s$-ensemble (see also~\cite{jack10-rom}).  
On taking an initial configuration from a metastable state $\alpha$ and simulating equilibrium dynamics, 
the probability that the system remains in state $\alpha$ throughout a long time $\tobs$ is $P(\alpha\to\alpha)\sim \ee^{-\gamma\tobs}$,
where $\gamma$ is a rate for relaxation to equilibrium.  For metastable states with long lifetimes, one expects a nucleation
mechanism for relaxation: nucleation may take place at any position in a large system so that $\gamma\propto N$ on
taking the thermodynamic limit $N\to\infty$.
Thus, for large enough $N$ and $\tobs$, one expects $P(\alpha\to\alpha)/P(\alpha\to\mathrm{eq}) \sim \ee^{-\gamma_0 N\tobs}$ 
where $P(\alpha\to\mathrm{eq})\approx1$ is the probability of the system relaxing back to equilibrium.

Let the mean dynamical activity $K$ for long trajectories localised in state $\alpha$ be $k_\alpha N\tobs$,
and the mean activity for trajectories that relax to equilibrium be $k_{\rm eq} N\tobs$.  
  Then, in the $s$-ensemble, Eq.~(\ref{equ:s-ens}) yields the ratio of probabilities for
remaining localised in state $\alpha$ and for relaxation to equilibrium,
\begin{equation}
\frac{P_s(\alpha\to\alpha)}{P_s(\alpha\to\mathrm{eq})} \sim \ee^{[s(k_\mathrm{eq}-k_\alpha)-\gamma_0] N\tobs}.
\label{equ:relax}
\end{equation}
where we assumed that $k_\alpha$ and $k_\mathrm{eq}$
depend only weakly on $s$ for small $s$, consistent with our observation that fast (vibrational, intra-state) degrees of freedom 
are affected weakly by $s$.
Eq. (\ref{equ:relax}) shows that if state $\alpha$ is less active than the equilibrium state ($k_\alpha<k_{\rm eq}$) 
and if $s>s^*=\gamma_0/(k_\mathrm{eq}-k_\alpha)$, 
then trajectories starting
in state $\alpha$ will remain localised in that state, and will not relax to equilibrium even as $\tobs\to\infty$. 
  This construction shows how metastable states that are irrelevant
at equilibrium may dominate the $s$-ensemble defined in (\ref{equ:s-ens}).  
[The probability of relaxation to a new metastable state $\alpha'\neq\alpha$ might
be larger than $P(\alpha\to\alpha)$ but
that is not relevant for the current argument.] 

The field $s^*$ required to
stabilise state $\alpha$ may
be very small if the metastable state is long-lived ($\gamma_0$ is small).
However, for small enough $s$, there is always a regime $s<s^*$, where
relaxation to equilibrium is preferred to localisation in a metastable state, 
as long as $\gamma_0$, $k_\alpha$ and $k_\mathrm{eq}$ are strictly positive (non-zero) constants.
The definition of $K$ considered here
ensures that $k_\alpha$ and $k_\mathrm{eq}$ are both finite.
Assuming finite short-ranged interaction potentials and
that the equilibrium state of the system is indeed a fluid,
the nucleation rate
$\gamma_0$ must also be non-zero even in the thermodynamic limit~\cite{KL11}.
Thus, for these systems, we expect any transitions
in the $s$-ensemble to take place at $s=s^*$, with $s^*$ strictly greater than zero.  There are exceptions
to this rule in idealised model systems: for example, in 
mean-field models it may be that $\gamma_0\to0$ as $N\to\infty$ due to diverging free energy barriers~\cite{jack10-rom}, while
``kinetic constraints'' can lead to $\gamma_0\to0$ in the thermodynamic limit~\cite{garrahan-fred}.  
Transitions at $s^*=0$ might also be possible if the difference in  activity
density $k_\mathrm{eq}-k_\alpha$ were to diverge, which may be relevant for glass formers \cite{fred}.

Finally, we note that in any system with long-lived metastable states, 
the ``mean-field'' analysis leading to Eq.~(\ref{equ:relax}) predicts a dynamic phase transition  
at $s=s^*$.  However, fluctuations may destroy these transitions.
For example, as well as $P(\alpha\to\alpha)$ and $P(\alpha\to\mathrm{eq})$, one should consider the possibility that
one part of a trajectory remains localised in state $\alpha$ while another part has a structure compatible with 
thermal equilibrium.  If this is likely, increasing $s$ may result in a smooth crossover from active to inactive behaviour,
with no dynamical phase transition.
As demonstrated in~\cite{elmatad10-pnas} for a kinetically constrained model, 
it is the strength of the coupling between the dynamics in different parts of a system
that determines whether a dynamical phase transition takes place.%

We conclude that the $s$-ensemble provides a most effective method for sampling metastable
states in glassy systems.  
By biasing trajectories according to their dynamical activity, the method samples these states ``democratically'',
without any assumptions about their structural features or long-ranged correlations.
In the KA mixture, we find metastable states that are associated with deep minima of the energy landscape and have few
soft vibrational modes.  
Now that these states can be prepared and characterised precisely, it will be interesting
to see whether their properties can be predicted and explained by theories of the glass transition.

We are grateful to N. Wilding, J. Kurchan, C. P. Royall and F. van Wijland for discussions.  This work was supported in part by 
EPSRC Grant no. EP/I003797/1 (to RLJ).  
In the early stages of this work LOH and DC were supported by 
NSF Grant No.\ CHE-0624807 and in its final stages by DOE Contract No.\ DE-AC0205CH11231.


\begin{thebibliography}{99}

\bibitem{ediger96}
M. Ediger, C. Angell, and S. Nagel, J. Phys. Chem. {\bf 100}, 13200 (1996).

\bibitem{deb-still01}
P. Debenedetti and F. Stillinger, Nature {\bf 410}, 259 (2001). 

\bibitem{cavagna}
A. Cavagna, Phys. Rep. {\bf 476}, 51 (2009).

\bibitem{ARPC}
D. Chandler and J.~P. Garrahan, Annu. Rev. Phys. Chem. {\bf 61}, 191 (2010).

\bibitem{berthier-review}
L. Berthier and G. Biroli, Rev. Mod. Phys. {\bf 83}, 587 (2011).

\bibitem{merolle-jack} M. Merolle, J.P. Garrahan and D. Chandler,
Proc. Natl. Acad. Sci. USA {\bf 102}, 10837 (2005); 
R.~L. Jack, J.~P. Garrahan and D. Chandler, J. Chem. Phys. {\bf 125}, 184509
(2006).

\bibitem{lecomte07}
V. Lecomte, C. Appert-Rolland, and F. van Wijland, J. Stat. Phys. {\bf 127}, 51 (2007). 

\bibitem{garrahan-fred}
J.~P. Garrahan, R.~L. Jack, V. Lecomte, E. Pitard, K. van Duijvendijk and F. van Wijland, 
Phys. Rev. Lett. {\bf 98}, 195702 (2007); J. Phys. A {\bf 42}, 075007 (2009). 

\bibitem{hedges09}
L.~O. Hedges, R.~L. Jack, J.P. Garrahan and D. Chandler, Science {\bf 323}, 1309 (2009).

\bibitem{jack10-rom}
R. L. Jack and J. P. Garrahan, Phys. Rev. E {\bf 81}, 011111 (2010).

\bibitem{elmatad10-pnas}
Y. S. Elmatad, R. L. Jack, D. Chandler, and J. P. Garrahan, Proc. Natl. Acad. Sci. USA {\bf 107}, 12793 (2010). 

\bibitem{Kurchan96}
J. Kurchan and L. Laloux, J. Phys. A {\bf 29}, 1929 (1996).

\bibitem{Heuer08}
A. Heuer, J. Phys: Cond. Matter {\bf20}, 373101 (2008)

\bibitem{tap}
D.~J.~Thouless, P.~W.~Anderson and R.~G.~Palmer, Phil. Mag. {\bf 35}, 593 (1977).

\bibitem{BK01}
G. Biroli and J. Kurchan, Phys. Rev. E {\bf64}, 016101 (2001).

\bibitem{heuer03}
B.~Doliwa and A. Heuer, Phys. Rev. Lett. {\bf 91}, 235501 (2003).

\bibitem{pts08}
G. Biroli, J.-P. Bouchaud, A. Cavagna, T.~S.~Grigera and P. Verocchio, Nature Physics {\bf 4}, 771 (2008).

\bibitem{KL11}
J. Kurchan and D. Levine, J. Phys. A {\bf 44} 035001 (2011).

\bibitem{ka95a}
W. Kob and H. C. Andersen, Phys. Rev. E {\bf 51}, 4626 (1995). 

\bibitem{berthier-kob07}
L. Berthier and W. Kob, J. Phys.: Condens. Matt. {\bf 19}, 205130 (2007). 

\bibitem{gc02-prl}
J.~P. Garrahan and D. Chandler,
Phys. Rev. Lett. {\bf 89}, 035704 (2002).

\bibitem{tps-annrev02}
P. Bolhuis, D. Chandler, C. Dellago, and P. Geissler, Ann. Rev. Phys. Chem. {\bf 53}, 291 (2002). 

\bibitem{ediger07}
S.~F.~Swallen, K.~L.~Kearns, M.~K.~Mapes, Y.~S.~Kim, R.~J.~McMahon, M.~D.~Ediger, T.~Wu, L.~Wi and S.~Satija, Science {\bf 315}, 353 (2007).

\bibitem{harrowell}
A.~Widmer-Cooper, H.~Perry, P.~Harrowell and D.~R.~Reichman, Nature Physics {\bf 4}, 711 (2008).

\bibitem{wyart07}
C. Brito and M. Wyart, J. Stat. Mech. (2007) L08003.

\bibitem{still-web84}
F. H. Stillinger and T. A. Weber, Science {\bf 225}, 983 (1984). 

\bibitem{saddles}
L.~Angelani, R.~Di Leonardo, G.~Ruocco, A.~Scala, and F.~Sciortino,
Phys. Rev. Lett. {\bf 85}, 5356 (2000);
T.~S.~Grigera, A.~Cavagna, I.~Giardina and G.~Parisi,
Phys. Rev. Lett. {\bf 88}, 055502 (2002);
D.~Coslovich and G.~Pastore, Europhys. Lett. {\bf 75}, 7840 (2006).

\bibitem{berthier2004}
L. Berthier, Phys. Rev. E {\bf 69}, 020201(R) (2004). 

\bibitem{laird}
S.~D.~Bembenek and B.~B~Laird, Phys. Rev. Lett. {\bf 74}, 936 (1995).

\bibitem{honey-cna}
J. D. Honeycutt and H. C. Andersen, J. Phys. Chem. {\bf 91}, 4950 (1987). 

\bibitem{ka-cna}
H.~Jonsson and H.~C.~Andersen, Phys. Rev. Lett. {\bf 60}, 2295 (1988).

\bibitem{fred}
E. Pitard, V. Lecomte and F. van Wijland,  arXiv:1105.2460.

\end{thebibliography}
\end{document}